\documentclass[a4paper]{article}
\usepackage{bbm}
\usepackage{amsmath,amsthm,amsfonts,amssymb,color,hyperref,anysize,enumitem,graphicx,epstopdf,algorithm,algpseudocode,cleveref,multirow}
\usepackage[usenames,dvipsnames]{xcolor}
\usepackage[margin=0.9in]{geometry}

\DeclareMathOperator*{\argmax}{arg\,max}

\newcommand{\calM}{\mathcal{M}}

\newcommand{\bbe}{\mathbf{e}}

\newcommand{\bbp}{\mathbf{p}}

\newcommand{\bbx}{\mathbf{x}}

\newcommand{\ep}{\epsilon}
\newcommand{\rr}{\mathbb{R}}

\newcommand{\la}{\leftarrow}
\newcommand{\ra}{\rightarrow}

\newcommand{\pc}{\bbp^\circ}

\newcommand{\ps}{p^*}

\newcommand{\tp}{\tilde{p}}

\newcommand{\tP}{\tilde{P}}
\newcommand{\comm}[1]{\qquad\mbox{(#1)}}

\newcommand{\G}{\Gamma}
\newcommand{\pt}{x^t}
\newcommand{\ptone}{x^{t-1}}

\newcommand{\pj}{p_j}

\newcommand{\Dp}{\Delta p}

\newcommand{\Lj}{L_{j}}
\newcommand{\Ljj}{L_{jj}}

\newcommand{\Ljk}{L_{jk}}
\newcommand{\Dt}{\Delta t}

\newcommand{\kt}{k_t}
\newcommand{\ktau}{k_\tau}
\newcommand{\bG}{\overline{\G}}

\newcommand{\Gjt}{\Gamma_j^t}

\newcommand{\Dpktt}{\Delta p_{k_t}^t}

\newcommand{\gktt}{g_{k_t,2}^t}

\newcommand{\pkt}{x_{k_t}}

\newcommand{\pktt}{x_{k_t}^t}

\newcommand{\pkttm}{x_{k_t}^{t-1}}
\newcommand{\pkttau}{x_{k_t}^{\tau}}

\newcommand{\tz}{\tilde{z}}

\renewcommand{\O}{\mathcal{O}}

\def\pjq{p_{j,q}}
\def\tzjq{\tilde z_{j,q}}
\def\D{\Delta}

\newtheorem{theorem}{Theorem}
\newtheorem{lemma}{Lemma}

\newtheorem{corollary}[lemma]{Corollary}

\newenvironment{pf}{\begin{proof}[\emph{\textbf{Proof: }}]}{\end{proof}}
\newenvironment{pfof}[1]{\begin{proof}[\emph{\textbf{Proof of #1: }}]}{\end{proof}}

\newcommand{\PPAD}{\textsf{PPAD}}
\newcommand{\FP}{\textsf{FP}}

\newcommand{\rrplusn}{\mathbb{R}_+^n}
\newcommand{\bps}{\mathbf{p}^*}

\newcommand{\rhoi}{{\rho_i}}

\newcommand{\xij}{x_{ij}}

\newcommand{\aij}{a_{ij}}

\newcommand{\hsp}{~~~~~~}

\title{Amortized Analysis of Asynchronous Price Dynamics}

\author{Yun Kuen Cheung\thanks{Part of the work done while this author was at the Courant Institute, NYU
		and at the Faculty of Computer Science, University of Vienna.
		He was supported in part by NSF Grant CCF-1217989, and the Vienna Science and Technology Fund (WWTF) project ICT10-002.
		Additionally the research leading to these results has received funding from
		the European Research Council under the European Union's Seventh Framework Programme (FP7/2007-2013) / ERC Grant Agreement no.~340506.}\\
	Max Planck Institute for Informatics\\Saarland Informatics Campus
\and
	Richard Cole\thanks{{This work was supported in part by NSF grant CCF-1527568.}}\\Courant Institute, NYU
}

\date{}

\begin{document}

\maketitle

\begin{abstract}
We extend a recently developed framework for analyzing asynchronous coordinate descent algorithms
to show that an asynchronous version of tatonnement,
a fundamental price dynamic widely studied in general equilibrium theory,
converges toward a market equilibrium for Fisher markets with CES utilities or Leontief utilities,
for which tatonnement is equivalent to coordinate descent.

\smallskip

\end{abstract}
\section{Introduction}\label{sect:intro}

As is well known, it is \PPAD-hard to compute equilibria for
general games and markets~\cite{DaskalakisGP09,ChenDT2009,CSVY2006,CDDT2009,VY2011,CPY2017}.
By viewing the players and the environment collectively as implicitly performing a computation, 
these hardness results indicate that, in general, a game or market cannot reach an equilibrium quickly
(assuming no unexpected complexity results such as $\PPAD = \FP$).
As a result, a lot of attention has been given to the design of polynomial-time algorithms to compute equilibria,
either exactly or approximately, for specific families of games and markets.
Most of these algorithms can be categorized as either simplex-like (e.g., Lemke-Howson~\cite{LH1964}), numerical methods
(e.g., the interior-point method~\cite{Ye2008} or the ellipsoid method~\cite{Jain04}),
or some carefully-crafted combinatorial algorithms
(e.g., flow-based algorithms for computing an equilibrium of a market with agents
having linear utility functions~\cite{DPSV08,Orlin10,DM2015}).

However, it seems implausible that these algorithms describe the implicit computations in games or markets.
In particular, many markets appear to have a highly distributed environment. 
This would appear to preclude computations which require centralized coordination,
which is essential for the three categories of algorithms above.
Consequently, in order to justify equilibrium concepts, we want natural algorithms
which could plausibly be running (in an implicit form) in the associated distributed environments.
Moreover, since it is preferable not to assume centralized timing or coordination,
a desirable feature of such natural algorithms is robustness against asynchrony,
which means such algorithms should remain effective even in situations where
information transfer takes time and agents make decisions (i.e., perform computations) with possibly outdated information.

A first candidate for a natural algorithm in markets is tatonnement:
it adjusts the price of a good upward if there is too much demand, and downward if too little.
Indeed, tatonnement was proposed alongside the concept of a market equilibrium by Walras \cite{Walras1874} in 1874.
Since then, studies of market equilibria and tatonnement have received much attention in economics, operations research,
and most recently in computer science;
we list a small sample of the voluminous literature, focusing mainly on computer science works~\cite{ABH1959,Uzawa1960,Dohtani1993,CMV2005,CF2008,CFR2010,CCR2012,CCD2013,PY2010}.
Underlying many of these works is the issue of what are plausible price adjustment mechanisms and in what types of markets they attain a market equilibrium.

The tatonnements studied in prior work have mostly been continuous, or discrete and synchronous.
Cole and Fleischer~\cite{CF2008} observed that real-world market dynamics are highly distributed and hence presumably asynchronous.
They argued that any realistic price dynamics must involve out-of-equilibrium trade in order to induce the imbalances leading to price updates.
Further, they argued that simple rules with relatively low information requirements were more plausible.
The lowest imaginable level of information would be for each seller to only know the demand for the good it was selling,
and for any price updating to occur in a non-coordinated manner, i.e., asynchronously.
Accordingly, they introduced the \emph{Ongoing market model}, 
a model of a repeating market incorporating update dynamics,
and they analyzed the performance of an asynchronous tatonnement in this market.
The market also incorporated warehouses (buffers) to cope with supply and demand imbalances.

Cheung, Cole and Devanur~\cite{CCD2013} showed that tatonnement is equivalent to coordinate descent
on a convex function for several classes of Fisher markets, and consequently
that a suitable synchronous tatonnement converges toward the market equilibrium in three general classes of markets:
complementary-CES Fisher markets\footnote{i.e., markets in which the buyers all have complementary CES utilities.},
substitute-CES Fisher markets, and Leontief Fisher markets;
Cheung~\cite{Cheung2014} extended this to all nested-CES Fisher markets.
In this paper, we show that this equivalence enables us to perform an amortized analysis
to show that the corresponding asynchronous version of tatonnement converges toward the market equilibrium in these classes of markets;
indeed, our analysis also covers Fisher markets in which some buyers have substitute-CES utility functions and others have complementary ones.
We also note that the tatonnement for Leontief Fisher markets analyzed in~\cite{CCD2013}
had an unnatural constraint on the step sizes; our analysis removes that constraint.

Finally, we remark that it is by no means obvious that the existence of a convergence result
for synchronous updating implies an analogous result for asynchronous updating.
An example of a setting where an asynchronous result has yet to be achieved
is proportional response dynamics~\cite{Zhang2011,BDX2011,CCT2018-EC}.

\paragraph{Technique of Analysis, and Comparison with the Companion Paper~\cite{CCT2018}.}
In a companion paper~\cite{CCT2018}, we analyzed several versions of asynchronous coordinate descent.
The analyses in both papers follow a common framework.
We use an amortized analysis which relates the actual progress to the desired progress,
where the desired progress is a constant fraction of the progress achieved with synchronous updating.
The amortization is used to hide the difference between these two measures of progress by amortizing it over multiple updates.
As we shall see, this difference is bounded by the squares of appropriate excess demand (resp.~gradient) differences,
and using Lipschitz gradient parameters, these can in turn be bounded by sum of the squares of recent changes to the prices (resp.~coordinates).
The final ingredient is to show that the progress is an upper bound on the square of the change to the updated price.
Combining these ingredients yields a lower bound on the rate of progress.

In~\cite{CCT2018}, it was assumed that the underlying convex function has some \emph{global} finite Lipschitz gradient parameters,
which is a common assumption in optimization and machine learning.
The main focus there is on the maximum possible degree of parallelism which permits linear speedup,
and on a number of challenges to devising rigorous and complete analyses
which handle the subtle interplay between randomness (choices of coordinates) and asynchrony.

However, in the asynchronous tatonnement setting we analyze here, there are no global finite Lipschitz gradient parameters.
Instead, we use \emph{local} Lipschitz gradient parameters, as was done implicitly in~\cite{CCD2013};
the consequence is that the rate of convergence depends on the starting point.
Also, the only acceptable degree of parallelism is the maximal one, i.e., all sellers are adjusting prices independently in parallel.
The challenge is to devise an asynchronous analysis while keeping the price update rule reasonable,
i.e., having the step size be an absolute constant which is independent of the number of goods.
This calls for a somewhat different potential function and analysis from the one used for the asynchronous
coordinate descent analysis in~\cite{CCT2018};
the analysis also differs quite substantially from the synchronous tatonnement analyses in~\cite{CCD2013}.

\paragraph{Relevance to Theoretical Computer Science.}
Iterative procedures and dynamical systems are pervasive across multiple disciplines;
a non-exhaustive list of such systems which have interested theorists includes
bandwidth sharing (e.g., proportional response~\cite{WZ2007,Zhang2011,CCT2018-EC}),
SDD linear system solvers~\cite{KOSZ2013,LS2013}, distributed load balancing~\cite{EKM2007,BFGGHM2007},
bird flocking~\cite{Chazelle2009}, influence systems~\cite{Chazelle2012}
and the spread of information memes across the Internet~\cite{LBK2009}.

There have been many analyses of these systems, but one issue that has received relatively little attention is the timing of agents' actions.
In most prior analyses, amenable timing schemes (e.g., synchronous or round robin updates) and perfect information retrieval were assumed,
perhaps because they were more readily analyzed. However, typically these assumptions are unrealistic,
and to better understand how these systems really behave, it is important to obtain asynchronous analyses of such systems. 
We believe the insight from our amortization framework may be useful in obtaining such analyses.

\paragraph{Other Related Work.}
In a similar spirit to our analysis, Cheung, Cole and Rastogi~\cite{CCR2012} analyzed asynchronous tatonnement in certain Fisher markets.
This earlier work employed a potential function which drops continuously when there is no update and does not increase when an update is made.
This approach could be followed for the current market setting, but in the current work,
we instead use a discrete analysis which has more in common with our asynchronous coordinate descent analyses in~\cite{CCT2018}.
Our work differs from~\cite{CCR2012} in two aspects.
First, the update rule in~\cite{CCR2012} is more restricted: they use \emph{average excess demand} for updates,
while our update rule allows an \emph{arbitrary value between the maximum and minimum excess demands}.
Second, while the high-level idea is similar, our potential function is substantially different from
(and more sophisticated than) the one in~\cite{CCR2012}, and the classes of markets covered by the two analyses are quite different.

In a recent work, Dvijotham et al.~\cite{DRS2016} study a different asynchronous dynamics.
In their setting sellers are boundedly rational and buyers are myopic (i.e., best responding).
More specifically, the base (zero) level for the sellers is to be best responding, and level
$k+1$ is obtained by best responding to level $k$ sellers.
They show that this system converges linearly to the market equilibrium in suitable Fisher markets including substitute-CES markets.

For the closely related topic of learning dynamics in games, where updates are based on the payoffs received by agents,
again, the classical approach assumes synchronous or round-robin updates with up-to-date payoffs;
models with stochastic update schedules were also studied previously (e.g., in~\cite{Blume1993,AN2010,MS2012}),
while learning dynamics with delayed payoffs~\cite{PalLa2015} were studied recently.
\section{Preliminaries and Results}\label{sect:prelim}

\paragraph{Fisher Market.}
In a Fisher market, there are $n$ perfectly divisible goods and $m$ buyers.
Without loss of generality, the supply of each good is normalized to be one unit.
Each buyer $i$ has a utility function $u_i:\rrplusn \ra \rr$, and a budget of size $e_i$.
At any given price vector $\bbp\in \rrplusn$, each buyer purchases a maximum utility
affordable collection of goods.
More precisely, $\bbx_i\in \rrplusn$ is said to be a demand of buyer $i$ if $~\bbx_i~\in~\argmax_{\bbx':~\bbx'\cdot \bbp \le e_i}~u_i(\bbx')$.

A price vector $\bps \in \rrplusn$ is called a \emph{market equilibrium} if at $\bps$, there exists a demand $\bbx_i$ of each buyer $i$ such that
$$
\ps_j > 0~~\Rightarrow~~\sum_{i=1}^m x_{ij} ~=~ 1
\hsp\hsp\text{and}\hsp\hsp
\ps_j = 0~~\Rightarrow~~\sum_{i=1}^m x_{ij} ~\le~ 1.
$$

We note that in the markets we studied here, the demand at any price vector is unique.
In these markets, we let $z_j ~:=~ \sum_{i=1}^m x_{ij} - 1$ denote the excess demand for good $j$.

\paragraph{CES utilities.}
In this paper, each buyer $i$'s utility function is of the form
$$
u_i(\bbx_i) ~=~ \left(\sum_{j=1}^n a_{ij} \cdot (x_{ij})^\rhoi\right)^{1/\rhoi},
$$
for some $-\infty \le \rhoi < 1$, where each $a_{ij}$ is a non-negative number.
$u_i(\bbx_i)$ is called a Constant Elasticity of Substitution (CES) utility function.
They are a class of utility functions often used in economic analysis.
The limit as $\rhoi \ra -\infty$ is called a Leontief utility, usually written as
$u_i(\bbx_i) = \min_j \frac{\xij}{c_{ij}}$ \footnote{The utility function $u_i(\bbx) = \min_j \frac{\xij}{c_{ij}}$ can be seen as the limit of $u_i(\bbx) = \Big( \sum_j \left(\frac{\xij}{c_{ij}}\right)^{\rho_i}\Big)^{\frac{1}{\rho_i}}$ as $\rho_i\searrow-\infty$.};
and the limit as $\rhoi \ra 0$ is called a Cobb-Douglas utility,
usually written as $\prod_j {\xij}^{a_{ij}}$, with $\sum_j \aij = 1$. 
The utilities with $\rhoi\le 0$ capture goods that are complements, and those with $\rhoi \ge 0$ goods that are substitutes.
Accordingly, when $\rhoi\le 0$, we say the utility function is a complementary CES utility function,
and when $\rhoi\ge 0$ we say it is a substitute CES utility function.

\paragraph{Directly Related Prior Results and Our Results.}
Cheung, Cole and Devanur~\cite{CCD2013} showed that tatonnement is equivalent to coordinate descent on a convex function $\phi$
for Fisher markets with buyers having complementary-CES or Leontief utility functions
(and in a later version of the paper, substitute-CES utility functions too).
To be specific, \cite{CCD2013} showed that for the convex function
$$\phi(\bbp) = \sum_{k=1}^n p_k ~+~ \sum_{i=1}^m e_i\cdot \log \hat{u}_i(\bbp),$$
where $\hat{u}_i(\bbp)$ is the optimal utility that buyer $i$ attains at price vector $\bbp$ with a unit of spending,
we have that $\nabla_j \phi(\bbp) = -z_j(\bbp)$.
The corresponding update rule is
\begin{equation}\label{eq:sync-tat-CES-rule}
p_j' ~~\la~~ p_j\cdot \left[~1 + \lambda \cdot \min\{z_j,1\}~\right],
\end{equation}
where $\lambda > 0$ is a suitable constant. 
As the update rule is multiplicative, they assumed that the initial prices were positive.

As argued in~\cite{CF2008}, when the economic activity is occurring over time,
it is natural to base each price update for a good on the excess demand 
observed by its seller since the time of the last price update to her good 
(possibly weighted toward more recent sales).
This perceived excess demand can be written as the product of the length of the time interval with
an instantaneous excess demand at some specific time in this interval, 
which yields the following modification of update rule \eqref{eq:sync-tat-CES-rule}.
\begin{equation}\label{eq:async-tat-CES-rule}
p_j' ~~\la~~ p_j\cdot \left[~1 + \lambda \cdot \min\{\tilde{z}_j,1\}\cdot (t-\alpha_j(t))~\right],
\end{equation}
where $\alpha_j(t)$ denotes the time of the latest update to price $j$ \emph{strictly} before time $t$,
$\tilde{z}_j$ is a value between the minimum and maximum instantaneous excess demands during the time interval $(\alpha_j(t), t)$,
and $\lambda > 0$ is a suitable constant. We assume that $t - \alpha_j(t) \le 1$ for all $t\ge 0$ and for all goods $j$.

As we will see, having $\lambda \le 1/25.5$ suffices.
In comparison, in the synchronous version~\cite{CCD2013}, $\lambda \le 1/6$ suffices.
This implies that the step sizes of the asynchronous tatonnement can be kept at a constant fraction of those used in its synchronous counterpart.

\begin{theorem}\label{thm:CES-Fisher-cnvge}
For $\lambda \le 1/25.5$, asynchronous tatonnement price updates using rule \eqref{eq:async-tat-CES-rule} converge linearly toward
the market equilibrium in any complementary-CES market, and they converge in any Leontief Fisher market.
\end{theorem}

\begin{theorem}\label{thm:full-CES}
Let $\calM$ be a Fisher market in which buyers have CES utility functions.
Suppose that $\rho := \max_i \rho_i < 1$ and $\min_i \rho_i > -\infty$ in $\calM$.
Let $E := \max \left\{ 1/(1-\rho) ~,~ 1 \right\}$.
Then for $\lambda \le 1/(26E)$, asynchronous tatonnement price updates using rule \eqref{eq:async-tat-CES-rule} converge linearly toward
the market equilibrium.
\end{theorem}

In the main body of the paper, we focus on the result concerning complementary-CES Fisher markets.
The analysis for Theorem~\ref{thm:full-CES} is just a small modification of the complementary case,
and is deferred to Appendix~\ref{app:subs-CES}.
For the Leontief Fisher markets, while the first part of the analysis is identical to the complementary case,
this is not enough to demonstrate convergence, and to do so requires substantially more effort;
the full analysis is deferred to Appendix~\ref{app:leontief}.

In an earlier version of this paper~\cite{CC2014-arxiv} on arXiv,
we proved Theorem~\ref{thm:CES-Fisher-cnvge} (except that $\lambda$ was slightly larger) using a potential function
which decreases continuously over time, as was the case for the analyses in ~\cite{CFR2010, CCR2012} also.
We believe the current analysis is considerably simpler.
The main advantage of the prior analysis at this point is that we extended it to account
for the warehouses in the Ongoing market model, albeit with a quite non-trivial argument.
This seems possible with the potential function in the present paper too,
but we suspect it would be of interest to at most a few specialists.

\newcommand{\Dbbp}{\Delta \bbp}

\paragraph{Standard Notation in Coordinate Descent.}
Let $\bbe_j$ denote the unit vector along coordinate (in our context, price) $j$.
A function $F$ is $L$-Lipschitz-smooth if for any $\bbp,\Dbbp\in\rr^n$, $\|\nabla F(\bbp+\Dbbp) - \nabla F(\bbp)\| ~\le~ L\cdot\|\Dbbp\|$.
For any coordinates $j,k$, a function $F$ is $\Ljk$-Lipschitz-smooth if for any $\bbp\in\rr^n$ and $r\in\rr$,
$|\nabla_k F(\bbp+r\cdot \bbe_j) - \nabla_k F(\bbp)| ~\le~ \Ljk\cdot |r|$.
Also, as is standard, $\Lj$ denotes $\Ljj$.
\section{Key Ideas and Lemmas}\label{sect:key}

\newcommand{\alphaj}{\alpha_j}
\newcommand{\alphak}{\alpha_k}
\newcommand{\alphaks}{\alpha_{k_s}}
\newcommand{\alphakt}{\alpha_{k_t}}
\newcommand{\alphaktau}{\alpha_{k_{\tau}}}
\newcommand{\dt}{\delta t}
\newcommand{\dtau}{\delta t_{\tau}}
\newcommand{\Dtau}{\Delta  t_{\tau}}
\newcommand{\Dsig}{\Delta t_{\sigma}}
\newcommand{\dsig}{\delta t_{\sigma}}
\renewcommand{\gktt}{g_{k_t}^t}
\newcommand{\gksigsig}{g_{k_{\sigma}}^{\sigma}}
\newcommand{\gksigtau}{g_{k_{\sigma}}^{\tau}}
\newcommand{\gktaunu}{g_{k_{\tau}}^{\nu}}
\newcommand{\gktautau}{z_{k_{\tau}}^{\tau}}
\newcommand{\tgktt}{\tilde{g}_{k_t}^t}
\newcommand{\tgksigtau}{\tilde{g}_{k_{\sigma}}^{\tau}}
\newcommand{\tgktautau}{\tilde{z}_{k_{\tau}}^{\tau}}
\newcommand{\ks}{k_s}
\newcommand{\Gktt}{\G_{k_t}^t}
\newcommand{\Gktautau}{\G_{k_{\tau}}^{\tau}}
\newcommand{\Gksigtau}{\G_{k_{\sigma}}^{\tau}}
\newcommand{\Gktausig}{\G_{k_{\tau}}^{\sigma}}
\newcommand{\hWksig}{\widehat{W}_{k_{\sigma}}}
\newcommand{\Itau}{I_{\tau}}
\newcommand{\lktaut}{\ell_k(\tau,t)}
\newcommand{\lksigtau}{\ell_k(\sigma,\tau)}
\newcommand{\lksignutau}{\ell_k(\sigma,\nutau)}
\newcommand{\nutau}{\nu(\tau)}
\renewcommand{\ptone}{\bbp^{t-1}}
\renewcommand{\pt}{\bbp^{t}}
\newcommand{\ptau}{\bbp^{\tau}}
\newcommand{\pnutau}{\bbp^{\nu(\tau)}}
\newcommand{\pnut}{\bbp^{\nu(t)}}
\newcommand{\ptp}{\bbp^{t+1}}
\newcommand{\pkssig}{p_{k_s}^{\sigma}}
\newcommand{\pksigsig}{p_{k_\sigma}^{\sigma}}
\newcommand{\pkstau}{p_k^{\tau}}
\newcommand{\pkst}{p_k^t}
\renewcommand{\pkt}{p_{k_t}}
\renewcommand{\pktt}{p_{k_t}^t}
\renewcommand{\pkttm}{p_{k_t}^{t-}}
\renewcommand{\pkttau}{p_{k_t}^{\tau}}
\newcommand{\pktautau}{p_{k_{\tau}}^{\tau}}
\newcommand{\pktausig}{p_{k_{\tau}}^{\sigma}}
\newcommand{\pksigsigm}{p_{k_{\sigma}}^{\sigma-}}
\newcommand{\pktautaum}{p_{k_{\tau}}^{\tau-}}
\newcommand{\pti}{p^{t_i}}
\newcommand{\ptip}{p^{t_{i+1}}}
\newcommand{\Lksigkssig}{L_{k_{\sigma}k_s}^{[\sigma,\sigma+1]}}
\newcommand{\Lktaukstau}{L_{k_{\tau}k_s}^{[\tau,\tau+1]}}
\newcommand{\Lktkt}{L_{k_t,k}^{[t,t+1]}}
\newcommand{\Lktauktau}{L_{k_{\tau},k}^{[\tau,\tau+1]}}
\newcommand{\tprev}{t_{prev}}
\newcommand{\tmin}{t_{\min}}
\newcommand{\tsig}{t_{\sigma}}
\newcommand{\ttau}{t_{\tau}}
\renewcommand{\tt}{t_t}
\newcommand{\ti}{t_i}
\newcommand{\tip}{t_{i+1}}

\newcommand{\deltau}{\delta_\tau}
\newcommand{\delnu}{\delta_\nu}
\newcommand{\pktaunum}{p_{\ktau}^{\nu-}}
\newcommand{\gkmaxtau}{z_k^{\max,\tau}}
\newcommand{\gkmintau}{z_k^{\min,\tau}}
\newcommand{\gktaumaxtau}{z_{\ktau}^{\max,\tau}}
\newcommand{\gktaumintau}{z_{\ktau}^{\min,\tau}}
\newcommand{\numofupdatek}{U_k}

For simplicity, we assume that at any particular time $t$, there is at most one update to one good.
In general, since there is no coordination between price updates of different goods,
it is possible that the prices of two goods are updated at the same \emph{moment};
but by using any arbitrary tie-breaking (perturbation) rule, our analysis extends to such cases.

Recall update rule \eqref{eq:async-tat-CES-rule}. 
For the purposes of our analysis,
for each update we now need to know the elapsed time since the previous update to the same coordinate,
or since time $0$ if it is the first update to that coordinate; for the update at time $\tau$, we denote this by $\Dtau$.
As explained in~\cite{CF2008}, in the Ongoing market model, all the sellers
need to know is the size of their warehouse stock at the times of the current update and the previous update,
which seem to be very natural information.
We let $\alphaj(\tau)$ denote the time of the most recent update to $\pj$
\emph{strictly} before time $\tau$, or time $0$ if there is no previous update to this price.
We let $\alpha(\tau)$ denote the time of the most recent update to \emph{any} price
\emph{strictly} before time $\tau$, or time $0$ if there is no previous update to any price.
And we let $\deltau ~:=~ \tau - \alpha(\tau)$,
the elapsed time since the most recent previous update to any price.

Suppose there is an update at time $t$.
We let $p_{\kt}$ denote the price updated at time $t$, we let
$\pktt$ denote its updated value,
and $\pkttm$ its value right before this update;
note that $\pkttm \equiv p_{\kt}^{\alpha(t)} \equiv p_{\kt}^{t-\delta_t}$.
Let $\Dpktt := \pktt - \pkttm$.
Also, we let $\tz_{\kt}^t$ denote the value of the excess demand used
in the update to price $\pkt$ at time $t$.
Finally, we let $\Gktt := \max\{1,\tz_{\kt}^t\} / (\lambda \pkttm)$.
Then update rule~\eqref{eq:async-tat-CES-rule} can be rewritten in the following form:
\begin{equation}\label{eq:async-tat-CES-rule-alt}
p_{\kt}^t ~~\la~~ p_{\kt}^{t-} ~+~ \frac{1}{\Gktt} \cdot \tz_{\kt}^t \cdot \Dt.
\end{equation}

In our analysis, when we write $\sum_{\tau\in I}$, where $I$ is a time interval,
the summation is summing over all updates that occurred in time interval $I$.

Let $z_{k}^t$ be the instantaneous excess demand for good $k$ right before a price update at time $t$.
We note that $z_{k}^t = -\nabla_{k} \phi(\bbp^{\alpha(t)})$.
For each update $\tau$, let $\gkmaxtau$ and $\gkmintau$ denote
the maximum and minimum of accurate excess demand values of good $k$ in the time interval $(\alpha_{\ktau}(\tau),\tau)$.

We also need to define local Lipschitz parameters: $L_{jk}^{[\tau_a,\tau_b]}$ is an upper bound
on the Lipschitz gradient parameter $L_{jk}$ of the function $\phi$
within a rectangular hull of those prices which might appear in the time interval $[\tau_a,\tau_b]$.
Observe that in update rule~\eqref{eq:async-tat-CES-rule},
since $\left|\min\{\tz_j,1\}\right|\le 1$ always,
the above-mentioned rectangular hull is finitely bounded, and furthermore, it shrinks as $\lambda$ gets smaller.

We use the following three lemmas. Lemma~\ref{lem:discrete-improvement-of-F-with-time} is modified from a standard lemma in coordinate descent to accommodate
local Lipschitz gradient continuity. Lemma~\ref{lem:power-mean-ine} is a direct consequence of the Power-Mean inequality.
Lemma~\ref{lem:W-shift} is a simple algebra exercise. See Appendix~\ref{app:missing} for the missing proofs.

\begin{lemma}\label{lem:discrete-improvement-of-F-with-time}
Suppose there is an update to coordinate $\kt$ at time $t$ according to rule~\eqref{eq:async-tat-CES-rule},
and suppose that $\lambda \le 1/10$.
Recall that $\tz_{\kt}$ is the value used in applying rule~\eqref{eq:async-tat-CES-rule}.
Then
$$
\phi(\bbp^{\alpha(t)}) - \phi(\pt) ~~\ge~~ \frac{\Gktt}{4}\cdot \frac{ (\Delta p_{\kt}^t)^2} {\Dt} ~-~ \frac{1}{\Gktt}\cdot \left(z_{\kt}^t - \tz_{\kt}\right)^2\cdot |\Dt|.
$$
\end{lemma}

\begin{lemma}\label{lem:power-mean-ine}
Suppose that $w_1,w_2,\cdots,w_\ell$ and $y_1,y_2,\cdots,y_\ell$ are non-negative numbers. Then\\
$\left(\sum_{j=1}^\ell w_j y_j\right)^2 ~~\le~~ \left(\sum_{j=1}^\ell w_j\right)\left( \sum_{j=1}^\ell w_j \cdot (y_j)^2 \right)$.
\end{lemma}

\begin{lemma}\label{lem:W-shift}
In Lemma~\ref{lem:discrete-improvement-of-F-with-time}, suppose that $\tz'_{\kt}$ were used
instead of $\tz_{\kt}$ in update rule~\eqref{eq:async-tat-CES-rule-alt}, but with $\Gktt$ unchanged.
Let the new $\Dp_{\kt}$ value be $\Delta' p_{\kt}^{t}$. Then
$$
\Gktt\cdot \frac{ (\Delta p_{\kt}^t)^2} {\Dt}~~\ge~~ \frac {\Gktt}2\cdot \frac{ (\Delta' p_{\kt}^t)^2} {\Dt} ~-~ \frac{1}{\Gktt}\cdot (\tz_{\kt} - \tz'_{\kt})^2 \cdot \Dt.
$$
\end{lemma}

In the RHS of the inequality in Lemma~\ref{lem:discrete-improvement-of-F-with-time}, we call the first term,
$\frac{\Gktt}{4}\cdot \frac{ (\Delta p_{\kt}^t)^2} {\Dt}$, a progress term, and we call the second term,
$-\frac{1}{\Gktt}\cdot \left(z_{\kt}^t - \tz_{\kt}\right)^2\cdot |\Dt|$, an error term.
The progress term is cut into two halves. The first half will be used to demonstrate progress of the convergence,
while the second half will be saved to compensate for the error terms in subsequent updates.
Accordingly, we design a potential function $\Phi(t)$ of the form $\Phi(t) := \phi(\pt) + A(t)$, where $A(t)\ge 0$ for all $t$ and $A(0) = 0$.
We call $A(t)$ the \emph{amortization bank}; its purpose is to save portions of the progress terms for future compensations.

By showing that $\Phi(t)$ reduces by a constant fraction $\varepsilon$ in every $\O(1)$ time units,
we can deduce that $\phi(\pt) \le \Phi(t) \le \Phi(0) \cdot (1-\varepsilon)^{\Theta(t)} = \phi(\pc) \cdot (1-\varepsilon)^{\Theta(t)}$, as desired.

We define the function $A(t)$ as follows:
$$
A(t) = c_1 \sum_{\tau\in (t-1,t]} ~~ 
\sum_{k\neq \ktau}~
(1 + \mathbbm{1}[{\cal E}(k,\tau,t)]) \cdot
\Lktauktau \cdot \frac{p_k^\tau}{p_{\ktau}^{\tau-}} \cdot \frac {\left( \Delta \pktautau \right)^2} {\Delta \ttau},
$$
where $c_1>0$ is a constant we will determine later,
and ${\cal E}(k,\tau,t)$ denotes the event that price $k$ is \emph{not} updated during 
the time interval $(\tau,t]$.
\section{Analysis}\label{sect:analysis}

Suppose there is an update at time $t\ge 2$.
We let $t_a$ denote the time of the latest update \emph{strictly} before time $(t-2)$, if any; otherwise, we let $t_a=0$.
We let $t_b$ denote the time of the earliest update in the time interval $[t-1,t]$.
By Lemma~\ref{lem:discrete-improvement-of-F-with-time},
\begin{align}
&\sum_{\tau\in (t_a,t]} \left[\Phi(\alpha(\tau)) - \Phi(\tau)\right]\nonumber\\
&~\ge~ \sum_{\tau\in (t_a,t]} \left[
~\frac{\Gktautau}{4}\cdot \frac{(\Delta p_{\ktau}^\tau)^2}{\Dtau} ~-~\frac{1}{\Gktautau} (\gktautau - \tgktautau)^2\cdot \Dtau
~\right] ~+~ A(t_a) - A(t)\nonumber\\
&~\stackrel{(*)}{\ge}~ \sum_{\tau\in (t_a,t]} \left[
~\frac{\Gktautau}{4}\cdot \frac{(\Delta p_{\ktau}^\tau)^2}{\Dtau} ~-~\frac{1}{\Gktautau} (\gktaumaxtau - \gktaumintau)^2\cdot \Dtau
~\right] ~+~ A(t_a) - A(t).\label{eq:tat-target-first}
\end{align}
Inequality $(*)$ holds because in the tatonnement setting,
both the accurate excess demand $\gktautau$ and
the inaccurate excess demand $\tgktautau$
must lie between $\gktaumaxtau$ and $\gktaumintau$.

For any $\nu\in (\alpha_{\ktau}(\tau),\tau)$, let $\Delta' p_{\ktau}^\nu$ denote the $\Delta p_{\ktau}$ value
if there were an update at time $\nu$ to price $p_{\ktau}$ using the accurate excess demand $z_{\ktau}^\nu$.
By Lemma \ref{lem:W-shift}, for each $\tau\in (t_a,t]$,
\begin{align}
&\frac{\Gktautau}{8} \cdot \frac{(\Delta p_{\ktau}^\tau)^2}{\Dtau}\nonumber\\
&~\ge~ \sum_{\nu\in (\max\{t_a,\alpha_{\ktau}(\tau)\} , \tau ]} ~ \left[~
\frac{\Gktautau}{16} \cdot \frac{(\Delta' p_{\ktau}^\nu)^2}{\Dtau} \cdot  \frac{\delnu}{\Dtau} 
~-~ \frac{1}{8 \Gktautau} (z_{\ktau}^\nu - \tz_{\ktau}^\tau)^2 \cdot \delnu ~\right]\nonumber\\
&~\ge~ \frac{\Gktautau}{16} \left(\sum_{\nu\in (\max\{t_a,\alpha_{\ktau}(\tau)\} , \tau ]} \frac{(\Delta' p_{\ktau}^\nu)^2}{(\Dtau)^2} \cdot \delnu\right)
~-~ \frac{1}{8 \Gktautau} \cdot ( \gktaumaxtau - \gktaumintau )^2.\label{eq:tat-target-second}
\end{align}

For any $k$ and any time $\nu$, 
let $\beta_k(\nu)$ denote the time of the earliest update to price $k$ on or after time $\nu$.
Combining \eqref{eq:tat-target-first} and \eqref{eq:tat-target-second} yields
\begin{align*}
&\sum_{\tau\in (t_a,t]} \left[\Phi(\alpha(\tau)) - \Phi(\tau)\right]\\
&\ge~ \sum_{\tau\in (t_a,t]} ~ \sum_{\nu\in (\max\{t_a,\alpha_{\ktau}(\tau)\} , \tau ]}
\frac{\Gktautau}{16} \cdot \frac{(\Delta' p_{\ktau}^\nu)^2}{(\Dtau)^2} \cdot \delnu
~+~ \left[~\sum_{\tau\in (t_a,t]} ~ \frac{\Gktautau}{8} \cdot \frac{(\Delta p_{\ktau}^\tau)^2}{\Dtau} ~+~ A(t_a) - A(t)~\right]\\
&~~~~~~~~-~ \sum_{\tau\in (t_a,t]} ~ \frac{9}{8\Gktautau} \cdot ( \gktaumaxtau - \gktaumintau )^2\\
&\ge~ \frac{1}{16} \sum_{\nu \in (t_a,t_b]} \delnu \sum_{k=1}^n \frac{\G_k^{\beta_k(\nu)}\cdot (\Delta' p_{k}^\nu)^2}{(\Delta t_{\beta_k(\nu)})^2}
~-~ \sum_{\tau\in (t_a,t]} ~ \frac{9}{8\Gktautau} \cdot ( \gktaumaxtau - \gktaumintau )^2\\
&~~~~~~~~+~ \left[~\sum_{\tau\in (t_a,t]} ~ \frac{\Gktautau}{8} \cdot \frac{(\Delta p_{\ktau}^\tau)^2}{\Dtau} ~+~ A(t_a) - A(t)~\right]\\
&\ge~ \frac{1}{16} \sum_{\nu \in (t_a,t_b]} \delnu \left[~\sum_{k=1}^n \frac{\G_k^{\beta_k(\nu)}(\Delta' p_{k}^\nu)^2}{(\Delta t_{\beta_k(\nu)})^2}
~+~ c_2 \cdot A(\alpha(\nu))~\right]
~-~ \sum_{\tau\in (t_a,t]} ~ \frac{9}{8\Gktautau} \cdot ( \gktaumaxtau - \gktaumintau )^2\\
&~~~~~~~~+~ \left[~\sum_{\tau\in (t_a,t]} ~ \frac{\Gktautau}{8} \cdot \frac{(\Delta p_{\ktau}^\tau)^2}{\Dtau} ~+~ A(t_a) ~-~ A(t) ~-~ \frac{c_2}{16} \sum_{\nu \in (t_a,t_b]} \delnu \cdot A(\alpha(\nu))~\right],
\end{align*}
for some small constant $c_2>0$ we will determine later.

In \cite{CCD2013}, it was proved that the function $\phi$ is strongly convex
in any region bounded away from zero prices, and that the maximum $\G$ value throughout the tatonnement
is upper bounded by a finite constant which depends on the starting price $\bbp^\circ$.\footnote{Their
argument concerned the synchronous setting, but it can be reused without change for the asynchronous setting.}
We denote the finite upper bound on all $\G$'s by $\bG$, and the strong convexity parameter of $\phi$ by $\mu_\phi$,
which also depends on the starting prices.
We let $\varepsilon ~:=~ \mu_\phi / \bG$.
Then it is a standard fact in optimization that
$$
\sum_{k=1}^n \frac{\G_k^{\beta_k(\nu)}(\Delta' p_{k}^\nu)^2}{(\Delta t_{\beta_k(\nu)})^2}
~=~ \sum_{k=1}^n \frac{1}{\G_k^{\beta_k(\nu)}} \cdot (z_k^{\nu})^2
~\ge~ \sum_{k=1}^n \frac{1}{\bG} \cdot (z_k^{\nu})^2
~\ge~ \varepsilon \cdot \phi(\bbp^{\alpha(\nu)}).
$$
Setting $\varepsilon' =  \min\{\varepsilon,c_2\}/16$ yields
\begin{align}
&\sum_{\tau\in (t_a,t]} \left[\Phi(\alpha(\tau)) - \Phi(\tau)\right]\nonumber\\
&\ge~ \varepsilon' \sum_{\nu \in (t_a,t_b]} \delnu \cdot \Phi(\alpha(\nu))
~-~ \sum_{\tau\in (t_a,t]} ~ \frac{9}{8\Gktautau} \cdot ( \gktaumaxtau - \gktaumintau )^2\nonumber\\
&~~~~~~~~+~ \left[~\sum_{\tau\in (t_a,t]} ~ \frac{\Gktautau}{8} \cdot \frac{(\Delta p_{\ktau}^\tau)^2}{\Dtau} ~+~ A(t_a) ~-~ A(t) ~-~ \frac{c_2}{16} \sum_{\nu \in (t_a,t_b]} \delnu \cdot A(\alpha(\nu))~\right].\label{eq:tat-target-final}
\end{align}

In the subsections below, we will prove that for a suitable choice of the $\G$ parameters and $c_1,c_2$,
the final two terms of~\eqref{eq:tat-target-final}, in sum, are non-negative.
Also, we will show that $\Phi$ is decreasing over time. With these, the above inequality implies that
$$
\Phi(t_a) - \Phi(t) ~=~ \sum_{\tau\in (t_a,t]} \left[\Phi(\alpha(\tau)) - \Phi(\tau)\right]
~\ge~ \varepsilon' \sum_{\nu \in (t_a,t_b]} \delnu \cdot \Phi(\alpha(\nu)) ~\ge~ \varepsilon' \cdot \Phi(t),
$$
and hence $\Phi(t) ~\le~ \Phi(t_a) / (1+\varepsilon')$. By iterating this (note that $t> t_a\ge t-3$), we obtain
$$
\phi(\bbp^t) ~\le~ \Phi(t) ~\le~ (1+\varepsilon')^{-(t/3-1)}\cdot \Phi(0) ~=~ (1+\varepsilon')^{-(t/3-1)}\cdot \phi(\bbp^\circ),
$$
thus demonstrating linear convergence.

\newcommand{\Lktaukttautauo}{L_{\ktau,k_t}^{[\tau,\tau+1]}}

\subsection{$\Phi$ is a Decreasing Function}\label{sect:CES-H-decreasing}

For any time $\tau$ at which there is an update, 
by Lemma~\ref{lem:discrete-improvement-of-F-with-time} and the definition of $A$, we have
\begin{align*}
\Phi(\alpha(\tau)) - \Phi(\tau)
&~\ge~ \frac{\Gktautau}{4}\cdot \frac{(\Delta p_{\ktau}^\tau)^2}{\Dtau} ~-~\frac{1}{\Gktautau} (\gktautau - \tgktautau)^2\cdot \Dtau
~+~ c_1 \sum_{\nu\in (\alpha_{\ktau}(\tau),\tau)} L_{k_\nu \ktau}^{[\nu,\nu+1]}\cdot \frac{p_{\ktau}^{\tau-}}{p_{k_\nu}^{\nu-}}
\cdot \frac{\left(\Delta p_{k_\nu}^\nu\right)^2}{\Delta t_\nu}\\
&~~~~~~~-~ 2c_1 \sum_{k\neq \ktau} L_{\ktau,k}^{[\tau,\tau+1]} \cdot \frac{p_k^\tau}{p_{\ktau}^{\tau-}}
\cdot \frac{\left(\Delta p_{\ktau}^\tau\right)^2}{\Dtau}.
\end{align*}
Next,
\begin{align}
&\left( \gktaumaxtau - \gktaumintau \right)^2\nonumber\\
&~\le~ \left( \sum_{\nu\in (\alpha_{\ktau}(\tau),\tau)} L_{k_\nu,\ktau}^{[\nu,\tau]} \left| \Delta p_{k_\nu}^\nu \right| \right)^2\nonumber\\
&~=~ \left( \sum_{\nu\in (\alpha_{\ktau}(\tau),\tau)} \left(L_{k_\nu,\ktau}^{[\nu,\tau]} \cdot \Delta t_\nu \cdot \frac{p_{k_\nu}^{\nu-}}{p_{\ktau}^{\nu-}}\right)
\cdot \left(\frac{\left| \Delta p_{k_\nu}^\nu \right|}{\Delta t_\nu} \cdot \frac{p_{\ktau}^{\nu-}}{p_{k_\nu}^{\nu-}}\right) \right)^2\nonumber\\
&~\le~ \left( \sum_{\nu\in (\alpha_{\ktau}(\tau),\tau)} L_{k_\nu,\ktau}^{[\nu,\tau]} \cdot \Delta t_\nu
\cdot \frac{p_{k_\nu}^{\nu-}}{p_{\ktau}^{\nu-}} \right)
\left(
\sum_{\nu\in (\alpha_{\ktau}(\tau),\tau)} L_{k_\nu,\ktau}^{[\nu,\tau]} \cdot \frac{p_{\ktau}^{\nu-}}{p_{k_\nu}^{\nu-}}
\cdot \frac{\left( \Delta p_{k_\nu}^\nu \right)^2}{\Delta t_\nu}
\right)
\comm{by Lemma \ref{lem:power-mean-ine}}\nonumber\\
&~\le~ \left( \sum_{\nu\in (\alpha_{\ktau}(\tau),\tau)} L_{k_\nu,\ktau}^{[\alpha_{\ktau}(\tau),\tau]} \cdot \Delta t_\nu
\cdot \frac{p_{k_\nu}^{\nu-}}{p_{\ktau}^{\tau-}} \right)
\left(
\sum_{\nu\in (\alpha_{\ktau}(\tau),\tau)} L_{k_\nu,\ktau}^{[\nu,\nu+1]} \cdot \frac{p_{\ktau}^{\tau-}}{p_{k_\nu}^{\nu-}}
\cdot \frac{\left( \Delta p_{k_\nu}^\nu \right)^2}{\Delta t_\nu}
\right).\label{eq:tat-gradient-error}
\end{align}
Combining the above two equations and recalling that $\Delta t_\tau \le 1$ yields
\begin{align}
&\Phi(\alpha(\tau)) - \Phi(\tau)\nonumber\\
&~\ge~ \left( \frac {\Gktautau}4 - 2c_1 \sum_{k\neq \ktau} L_{\ktau,k}^{[\tau,\tau+1]} \cdot \frac{p_k^\tau}{p_{\ktau}^{\tau-}} \right)
\frac{\left(\Delta p_{\ktau}^\tau\right)^2}{\Dtau}\nonumber\\
&~~~~~+~ \left( c_1 - \frac{1}{\Gktautau} \sum_{\nu\in (\alpha_{\ktau}(\tau),\tau)} L_{k_\nu,\ktau}^{[\alpha_{\ktau}(\tau),\tau]} \cdot \Delta t_\nu
\cdot \frac{p_{k_\nu}^{\nu-}}{p_{\ktau}^{\tau-}} \right)\left(
\sum_{\nu\in (\alpha_{\ktau}(\tau),\tau)} L_{k_\nu,\ktau}^{[\nu,\nu+1]} \cdot \frac{p_{\ktau}^{\tau-}}{p_{k_\nu}^{\nu-}}
\cdot \frac{\left( \Delta p_{k_\nu}^\nu \right)^2}{\Delta t_\nu}\right).\label{eq:decreasing}
\end{align}

Thus, for $\Phi$ to be decreasing, we impose the following conditions (the second one is stronger than what is needed at this point):
\begin{equation}\label{eq:condition-one}
\Gktautau ~\ge~ 8c_1 \sum_{k\neq \ktau} L_{\ktau,k}^{[\tau,\tau+1]} \cdot \frac{p_k^\tau}{p_{\ktau}^{\tau-}}~~~\text{and}~~~
\Gktautau ~\ge~ \frac{2}{c_1} \sum_{\nu\in (\alpha_{\ktau}(\tau),\tau)} L_{k_\nu,\ktau}^{[\alpha_{\ktau}(\tau),\tau]} \cdot \Delta t_\nu
\cdot \frac{p_{k_\nu}^{\nu-}}{p_{\ktau}^{\tau-}}.
\end{equation}

\subsection{The Sum of the Last Two Terms in \eqref{eq:tat-target-final} is Non-negative}

It remains to show that the sum of the last two terms in \eqref{eq:tat-target-final} is non-negative, i.e.,
\begin{equation}
\sum_{\tau\in (t_a,t]} \frac{\Gktautau}{8}\cdot \frac{(\Delta p_{\ktau}^\tau)^2}{\Dtau} - \frac {c_2}{16} \sum_{\nu\in (t_a,t_b]} \delnu\cdot A(\alpha(\nu)) ~+~ A(t_a) - A(t)
~~\ge~~
\sum_{\tau\in (t_a,t]} \frac{9}{8\Gktautau} \cdot ( \gktaumaxtau - \gktaumintau )^2.\label{eqn::cond-to-satisfy}
\end{equation}
We first simplify the LHS using the definition of $A$:
\begin{align*}
&\sum_{\tau\in (t_a,t]} \frac{\Gktautau}{8}\cdot \frac{(\Delta p_{\ktau}^\tau)^2}{\Dtau}
~-~ \frac {c_2}{16} \sum_{\nu\in (t_a,t_b]} \delnu\cdot A(\alpha(\nu))
~+~ A(t_a) ~-~ A(t)\\
&~\ge~ \sum_{\tau\in (t_a,t]} \frac{\Gktautau}{8}\cdot \frac{(\Delta p_{\ktau}^\tau)^2}{\Dtau}
~-~ \frac {c_2}{16} ~\sum_{\nu\in (t_a-1,t_b]}~ 4c_1 \sum_{k\neq k_\nu}~ L_{k_\nu,k}^{[\nu,\nu+1]} \cdot \frac{p_k^{\nu}}{p_{k_\nu}^{\nu-}}
\cdot \frac{\left(\Delta p_{k_\nu}^\nu\right)^2}{\Delta t_\nu}\\
&\hspace*{1.5in}
+~ c_1 ~\sum_{\nu\in (t_a-1,t_a]}~ \sum_{k\neq k_\nu}~ L_{k_\nu,k}^{[\nu,\nu+1]} \cdot \frac{p_k^{\nu}}{p_{k_\nu}^{\nu-}}
\cdot \frac{\left(\Delta p_{k_\nu}^\nu\right)^2}{\Delta t_\nu}\\
&\hspace*{1.5in}
-~ 2c_1 ~\sum_{\nu\in (t-1,t]}~ \sum_{k\neq k_\nu}~ L_{k_\nu,k}^{[\nu,\nu+1]} \cdot \frac{p_k^{\nu}}{p_{k_\nu}^{\nu-}}
\cdot \frac{\left(\Delta p_{k_\nu}^\nu\right)^2}{\Delta t_\nu}\\
&~\ge~ \left( c_1 - \frac {c_1 c_2}{4} \right)
\sum_{\nu\in (t_a-1,t_a]}~ \sum_{k\neq k_\nu}~ L_{k_\nu,k}^{[\nu,\nu+1]} \cdot \frac{p_k^{\nu}}{p_{k_\nu}^{\nu-}}
\cdot \frac{\left(\Delta p_{k_\nu}^\nu\right)^2}{\Delta t_\nu}\\
&~~~~~~~~~~+~~
\sum_{\tau\in (t_a,t]}\left[\frac{\Gktautau}{8} - \left( \frac{c_1 c_2}{4} + 2c_1 \right) \sum_{k\neq k_\tau}~L_{\ktau,k}^{[\tau,\tau+1]} \cdot \frac{p_k^{\tau}}{p_{\ktau}^{\tau-}}\right]~\frac{(\Delta p_{\ktau}^\tau)^2}{\Dtau}.
\end{align*}

By imposing the requirement that
$\Gktautau ~\ge~ 8c_3 ~ \sum_{k\neq k_\tau}~L_{\ktau,k}^{[\tau,\tau+1]} \cdot \frac{p_k^{\tau}}{p_{\ktau}^{\tau-}}$,
for some constant $c_3>0$ which we will determine later, we obtain
\begin{align*}
&\sum_{\tau\in (t_a,t]} \frac{\Gktautau}{8} \frac{(\Delta p_{\ktau}^\tau)^2}{\Dtau}
~-~ \frac {c_2}{16} \sum_{\nu\in (t_a,t_b]} \delnu\cdot A(\alpha(\nu))
~+~ A(t_a) ~-~ A(t)\\
&~~\ge~~ \min \left\{~ c_1 - \frac{c_1 c_2}{4}~,~ c_3 - \frac{c_1 c_2}{4} - 2c_1 ~\right\} ~\cdot~
\sum_{\tau\in(t_a-1,t]} ~\sum_{k\neq \ktau}~ L_{k_\tau,k}^{[\tau,\tau+1]} \cdot \frac{p_k^{\tau}}{p_{\ktau}^{\tau-}}
~\frac{(\Delta p_{\ktau}^\tau)^2}{\Dtau}.
\end{align*}

On the other hand, by \eqref{eq:tat-gradient-error} and by the second condition imposed in~\eqref{eq:condition-one},
\begin{align*}
\sum_{\tau\in (t_a,t]} \frac{9}{8\Gktautau} \cdot ( \gktaumaxtau - \gktaumintau )^2
&~\le~ \frac{9c_1}{16} \sum_{\tau\in (t_a,t]}
~\sum_{\nu\in ( \alpha_{\ktau}(\tau) , \tau )}~ L_{k_\nu,\ktau}^{[\nu,\nu+1]}
\cdot \frac{p_{k_\tau}^{\tau-}}{p_{k_\nu}^{\nu-}} \cdot \frac {\left( \Delta p_{k_\nu}^\nu \right)^2} {\Delta t_\nu}\\
&~= ~ \frac{9c_1}{16} \sum_{\tau\in (t_a,t]}
 \sum_{\nu\in ( \alpha_{\ktau}(\tau) , \tau )}~ L_{k_\nu,\ktau}^{[\nu,\nu+1]}
 \cdot \frac{p_{k_\tau}^{\nu}}{p_{k_\nu}^{\nu-}} \cdot \frac {\left( \Delta p_{k_\nu}^\nu \right)^2} {\Delta t_\nu}\\
&~\le~ \frac {9c_1}{16} \sum_{\nu\in(t_a-1,t]} ~\sum_{k\neq k_\nu}~ L_{k_\nu,k}^{[\nu,\nu+1]} \cdot \frac{p_k^{\nu}}{p_{k_\nu}^{\nu-}}
~\frac{(\Delta p_{k_\nu}^\nu)^2}{\Dt_\nu}.
\end{align*}

By the above two inequalities, to satisfy~\eqref{eqn::cond-to-satisfy},
it suffices to have $\frac {9c_1}{16} ~\le~ \min \left\{~ c_1 - \frac{c_1 c_2}{4}~,\right.$
$\left. ~ c_3 - \frac{c_1 c_2}{4} - 2c_1 ~\right\}$.
There are multiple possible choices for $c_1,c_2,c_3$.
We choose $c_3 = 21c_1/8$ and $c_2 = 1/4$.
To summarize, we need the $\G$ parameters to satisfy
\begin{equation}\label{eq:Gamma-requirement}
\Gktautau ~\ge~ 21c_1 ~ \sum_{k\neq k_\tau}~L_{\ktau,k}^{[\tau,\tau+1]} \cdot \frac{p_k^{\tau}}{p_{\ktau}^{\tau-}} ~~~\text{and}~~~
\Gktautau ~\ge~ \frac{2}{c_1} \sum_{\nu\in (\alpha_{\ktau}(\tau),\tau)} L_{k_\nu,\ktau}^{[\alpha_{\ktau}(\tau),\tau]} \cdot \Delta t_\nu
\cdot \frac{p_{k_\nu}^{\nu-}}{p_{\ktau}^{\tau-}}.
\end{equation}
Our remaining tasks are to derive upper bounds on the two summations in~\eqref{eq:Gamma-requirement}.
\subsection{Upper Bounds on the Local Lipschitz Parameters, and Determining the $\G$'s}\label{sect:lip}

Suppose in a Fisher market with buyers having CES utility functions,
each buyer $i$ has a budget of $e_i$, and her CES utility function has parameter $\rho_i$.
For each $i$, let $\theta_i := \rho_i / (\rho_i - 1)$.
As we have discussed in Section~\ref{sect:prelim}, at any given price vector $\bbp\in \rrplusn$,
buyer $i$ computes the demand-maximizing bundle of goods costing at most $e_i$;
we let $x_{i\ell}(\bbp)$ denote buyer $i$'s demand for good $\ell$ at price vector $\bbp$.

In a Fisher market with buyers having complementary-CES utility functions,
the following properties are well-known. (See~\cite{CCR2012}.) 
\begin{enumerate}
\item For any $k\neq j$,
$$\left|\frac{\partial^2 \phi}{\partial p_j ~ \partial p_k}(\bbp) \right| ~=~ \sum_{i=1}^m \frac{\theta_i~x_{ij}(\bbp)~x_{ik}(\bbp)}{e_i} ~\leq~ \sum_{i=1}^m \frac{x_{ij}(\bbp)~x_{ik}(\bbp)}{e_i}.$$
\item Given positive price vector $\bbp$, for any $0 < r_1 < r_2$,
let $\bbp'$ be prices such that for all $\ell$, $r_1 p_\ell \leq p'_\ell \leq r_2 p_\ell$.
Then for all $\ell$, $\frac{1}{r_2} x_\ell(\bbp) \leq x_\ell(\bbp') \leq \frac{1}{r_1} x_\ell(\bbp)$.
\end{enumerate}

\begin{lemma}\label{lem:ces-bounds}
If the parameter $\lambda$ in update rule \eqref{eq:async-tat-CES-rule} is at most $1/10$, then
$$
\sum_{k\neq k_\tau} L_{k_\tau,k}^{[\tau,\tau+1]} \cdot \frac{p_k^{\tau}}{p_{\ktau}^{\tau-}} ~~\le~~ e^{4\lambda(\lambda+1)}\cdot \frac{x_{\ktau}(\bbp^{\tau-})}{p_{\ktau}^{\tau-}}
$$
and
$$
\sum_{\nu\in (\alpha_{\ktau}(\tau),\tau)} L_{k_\nu,\ktau}^{[\alpha_{\ktau}(\tau),\tau]} \cdot \Delta t_\nu
\cdot \frac{p_{k_\nu}^{\nu-}}{p_{\ktau}^{\tau-}}
~~\le~~ 2e^{8\lambda(\lambda+1)}\cdot \frac{x_{\ktau}(\bbp^{\tau-})}{p_{k_\tau}^{\tau-}}.
$$
\end{lemma}

\newcommand{\lamp}{\lambda'}

\begin{pf}
Let $\lamp = 2\lambda(\lambda+1)$.
Since $\lambda \leq 1/10$, it is easy to observe that for any $\nu\in [\tau,\tau+1]$
and for any $k$ (including coordinate $\ktau$),
\begin{equation}\label{eq:price-range}
e^{-\lamp}\cdot p_k^{\tau-} ~~\le~~ p_k^{\nu} ~~\le~~ e^{\lamp}\cdot p_k^{\tau-},
\end{equation}
on noting that the $\Delta t_{\nu}$ terms span up to 2 time units.

Accordingly, let
$\tP := \left\{(\tp_1,\tp_2,\cdots,\tp_n)\,\left|\,\forall k\in [n],~
e^{-\lamp} \cdot p_k^{\tau-} ~\le~ \tp_k ~\le~ e^{\lamp} \cdot p_k^{\tau-}\right.\right\}.$
Then,
\begin{align}
&\sum_{k\neq k_\tau} L_{k_\tau,k}^{[\tau,\tau+1]} \cdot \frac{p_k^{\tau}}{p_{\ktau}^{\tau-}}\nonumber\\
&~\le~ \frac{1}{p_{\ktau}^{\tau-}}~\sum_{k\neq \ktau} \left(\max_{\tp\in\tP}~\left|\frac{\partial^2 \phi}{\partial p_{\ktau} ~ \partial p_k} (\tp)\right|\right) \cdot p_k^{\tau}\nonumber\\
&~\le~ \frac{1}{p_{\ktau}^{\tau-}}~\sum_{k\neq \ktau} \sum_{i=1}^m~\frac{(e^{\lamp} x_{i\ktau}(\bbp^{\tau-}))\cdot(e^{\lamp} x_{ik}(\bbp^{\tau-}))}{e_i}
\cdot p_k^{\tau-}  \comm{by Properties 1 and 2}\label{eq:inter}\\
&~\le~ \frac{e^{2\lamp}}{p_{\ktau}^{\tau-}}~\sum_{i=1}^m~x_{i\ktau}(\bbp^{\tau-})~\sum_{k\neq k_\tau} \frac{x_{ik}(\bbp^{\tau-})\cdot p_k^{\tau-}}{e_i}\nonumber\\
&~\le~ \frac{e^{2\lamp}}{p_{\ktau}^{\tau-}}~\sum_{i=1}^m~x_{i\ktau}(\bbp^{\tau-})
\comm{the second summation is at most $1$, due to the budget constraint}\nonumber\\
&~=~ e^{2\lamp}\cdot \frac{x_{\ktau}(\bbp^{\tau-})}{p_{\ktau}^{\tau-}}.\label{eq:first-part}
\end{align}

For the time range $\nu\in [\alpha_{\ktau}(\tau),\tau]$, inequality \eqref{eq:price-range} also holds. Thus,

\begin{align*}
&\sum_{\nu\in (\alpha_{\ktau}(\tau),\tau)}~L_{k_\nu,\ktau}^{[\alpha_{\ktau}(\tau),\tau]} \cdot \Delta t_\nu
\cdot \frac{p_{k_\nu}^{\nu-}}{p_{\ktau}^{\tau-}}\\
&~\le~ \frac{1}{p_{\ktau}^{\tau-}}~\sum_{\nu\in (\alpha_{\ktau}(\tau),\tau)}~L_{k_\nu,\ktau}^{[\alpha_{\ktau}(\tau),\tau]} \cdot \Delta t_\nu
\cdot e^{2\lamp}\cdot p_{k_\nu}^{\tau-}\\
&~\le~ \frac{e^{2\lamp}}{p_{\ktau}^{\tau-}}~\sum_{k\neq \ktau}~L_{k,\ktau}^{[\alpha_{\ktau}(\tau),\tau]} \cdot p_k^{\tau-} \cdot
\sum_{\stackrel{\nu\in(\alpha_{\ktau}(\tau),\tau)}{k_\nu=k}} \Delta t_\nu\\
&~\le~ \frac{2e^{2\lamp}}{p_{\ktau}^{\tau-}}~\sum_{k\neq \ktau}~L_{k,\ktau}^{[\alpha_{\ktau}(\tau),\tau]} \cdot p_k^{\tau-}.~~~~~~\text{(observe that the $\sum_\nu \Delta t_\nu$ term above is at most $2$)}
\end{align*}
The summation $\sum_{k\neq \ktau}~L_{k,\ktau}^{[\alpha_{\ktau}(\tau),\tau]} \cdot p_k^{\tau-}$
above can be bounded as in \eqref{eq:first-part}, yielding an upper bound of $e^{2\lamp}\cdot x_{\ktau}(\bbp^{\tau-})$.
\end{pf}

To conclude, by \eqref{eq:Gamma-requirement} and Lemma \ref{lem:ces-bounds}, it suffices to have:
$$
\frac{1}{\lambda p_{\ktau}^{\tau-}} \cdot \max\{1,\tz_{\ktau}\}~=~ \Gktautau ~\ge~ \max \left\{ 21c_1 e^{4\lambda(\lambda+1)} ~,~ 
\frac{4}{c_1}\cdot e^{8\lambda(\lambda+1)} \right\} \cdot \frac{x_{\ktau}(\bbp^{\tau-})}{p_{\ktau}^{\tau-}},
$$
or equivalently,
$$
\lambda \cdot \max \left\{ 21c_1 e^{4\lambda(\lambda+1)} ~,~ 
\frac{4}{c_1}\cdot e^{8\lambda(\lambda+1)} \right\} ~~\le~~ \frac{\max\{1,\tz_{\ktau}\}}{x_{\ktau}(\bbp^{\tau-})}.
$$
The minimum possible value of the RHS above is $1/(2e^{2\lambda(\lambda+1)})$.
Thus, we need that
$$
\lambda \cdot \max \left\{ 42c_1 e^{6\lambda(\lambda+1)} ~,~ 
\frac{8}{c_1}\cdot e^{10\lambda(\lambda+1)} \right\} ~\le~ 1.
$$
We choose $c_1$ such that the two parameters in the $\max$ are equal, i.e., $c_1 = \frac{2}{\sqrt{21}}\cdot e^{2\lambda(\lambda+1)}$.
Then the above inequality reduces to $4\sqrt{21}\cdot \lambda \cdot e^{8\lambda(\lambda+1)} \le 1$; $\lambda \le 1/25.5$ suffices.

\section*{Acknowledgements}

We thank several anonymous reviewers for their helpful suggestions.

\bibliographystyle{plain}
\bibliography{bib}

\appendix

\newpage

\section{The Full Range of CES Utility Functions}\label{app:subs-CES}

To derive the upper bounds on the local Lipschitz parameters for substitute-CES utility functions,
we only need a few modifications from the complementary case.
Recall that $\theta_i := \rho_i / (\rho_i - 1)$.
Let $\rho$ denote the maximum $\rho_i$ among all buyers $i$, and let 
$$
\theta := \max\left\{\frac{\rho}{1-\rho}~,~1\right\}~~~~~~\text{and}~~~~~~E := \max \left\{ \frac{1}{1-\rho} ~,~ 1 \right\}.
$$
We note that when all $\rho_i$ are negative, then $\theta=E=1$.
The following facts are known (for Property 2, see~\cite{CF2008,CFR2010}):

\begin{enumerate}
\item For any $k\neq j$,
$$\left|\frac{\partial^2 \phi}{\partial p_j ~ \partial p_k}(\bbp) \right| ~=~ \left|\sum_{i=1}^m~ \frac{\theta_i~x_{ij}(\bbp)~x_{ik}(\bbp)}{e_i}\right|
~\le~ \theta \cdot \sum_{i=1}^m~\frac{x_{ij}(\bbp)~x_{ik}(\bbp)}{e_i}.$$
\item Given positive price vector $\bbp$, for any $1 \le r$,
let $\bbp'$ be prices such that for all $\ell$, $\frac{1}{r}\cdot p_\ell \leq p'_\ell \le r\cdot p_\ell$.
Then for all $\ell$,
$$r^{-(2E-1)} \cdot x_\ell(\bbp) \leq x_\ell(\bbp') \leq r^{2E-1} \cdot x_\ell(\bbp).$$
\end{enumerate}

Now we state the needed modifications in the proof of Lemma~\ref{lem:ces-bounds}.
First, both bounds will be multiplied by the factor $\theta$.
Second, in~\eqref{eq:inter}, we replace the two $e^{\lamp}$ by $e^{\lamp (2E-1)}$ in accord with the new Property 2.

\begin{lemma}\label{lem:subs-ces-bounds}
If the parameter $\lambda$ in update rule \eqref{eq:async-tat-CES-rule} is at most $1/10$, then
$$
\sum_{k\neq k_\tau} L_{k_\tau,k}^{[\tau,\tau+1]} \cdot \frac{p_k^{\tau}}{p_{\ktau}^{\tau-}} ~~\le~~ \theta e^{(8E-4)\lambda(\lambda+1)}\cdot \frac{x_{\ktau}(p^{\tau-})}{p_{\ktau}^{\tau-}}
$$
and
$$
\sum_{\nu\in (\alpha_{\ktau}(\tau),\tau)} L_{k_\nu,\ktau}^{[\alpha_{\ktau}(\tau),\tau]} \cdot \Delta t_\nu
\cdot \frac{p_{k_\nu}^{\nu-}}{p_{\ktau}^{\tau-}}
~~\le~~ 2\theta e^{8E\lambda(\lambda+1)}\cdot \frac{x_{\ktau}(p^{\tau-})}{p_{k_\tau}^{\tau-}}.
$$
\end{lemma}

To summarize, we need $\lambda \le 1/(10E)$, and
$$
\frac{1}{\lambda p_{\ktau}^{\tau-}} \cdot \max\{1,\tz_{\ktau}\}~=~ \Gktautau ~\ge~ \max \left\{ 21c_1\theta e^{(8E-4)\lambda(\lambda+1)} ~,~ 
\frac{4\theta}{c_1}\cdot e^{8E\lambda(\lambda+1)} \right\} \cdot \frac{x_{\ktau}(\bbp^{\tau-})}{p_{\ktau}^{\tau-}},
$$
or equivalently,
$$
\lambda\theta  \cdot \max \left\{ 21c_1 e^{(8E-4)\lambda(\lambda+1)} ~,~ 
\frac{4}{c_1}\cdot e^{8E\lambda(\lambda+1)} \right\} ~~\le~~ \frac{1}{2e^{2\lambda(\lambda+1)}}.
$$
We pick $c_1 = \frac{2}{\sqrt{21}}\cdot e^{2\lambda(\lambda+1)}$, then we need $4\sqrt{21} \cdot \lambda \theta \cdot e^{8E\lambda(\lambda+1)} \le 1$.
Observe that $\theta\le E$ always, so having $\lambda \le 1/(26E)$ suffices.

\section{Missing Proofs}\label{app:missing}

First of all, we need the following definition.
Let $C$ be a convex set and $F$ be a convex function.
The \emph{local Lipschitz gradient parameter} $L_j\ge 0$ within $C$ satisfies:
for all $\bbp\in C$ and $r\in \rr$ such that $\bbp + r\cdot \bbe_j \in C$,
$$\left| \nabla_j F(\bbp+r\cdot \bbe_j) - \nabla_j F(\bbp) \right| \le L_j \cdot |r|.$$
As is well-known, this is equivalent to:
$$
F(\bbp+r\cdot \bbe_j) ~-~ F(\bbp) ~-~ \nabla_j F(\bbp) \cdot r ~~\le~~ \frac{L_j}{2} \cdot r^2.
$$

We will prove a generalization of Lemma~\ref{lem:discrete-improvement-of-F-with-time}
which can be used to prove both Theorems~\ref{thm:CES-Fisher-cnvge} and~\ref{thm:full-CES}.
We need the following lemma from~\cite[Lemma 10.15]{Cheung2014}\footnote{Lemma~\ref{lem:upper-sandwich} is tailored to fit the scope of this paper.
\cite[Lemma 10.15]{Cheung2014} works for all nested-CES utility functions,
and it concerns some Bregman divergences which are not needed here. The inequality stated in Lemma~\ref{lem:upper-sandwich}
comes from the next to last line of the proof of \cite[Lemma 10.15]{Cheung2014}.};
a weaker version of this lemma which concerns complementary-CES Fisher markets
or substitute-CES Fisher markets can be found in~\cite{CCD2013}.

\begin{lemma}\label{lem:upper-sandwich}
In the Fisher market $\calM$ described in Theorem~\ref{thm:full-CES},
if for each $j$, $\left|\frac{\Dp_j}{p_j}\right| \le \min \left\{ 1/4~,~1/E \right\}$, then
$$\phi(\bbp+\Delta \bbp) - \phi(\bbp) + \sum_{j=1}^n z_j \cdot \Dp_j ~\le~ \left( \frac 23 + \frac{21E}{25} \right)\sum_{j=1}^n \frac{x_j}{p_j} (\Dp_j)^2.$$
In other words, within the convex set in which the price of every good $k\neq j$ is same as $p_k$,
and the price of good $j$ is within a factor $\left( 1\pm \left\{ 1/4, 1/E \right\} \right)$ of $p_j$,
the local Lipschitz gradient parameter $L_j$ is at most $\left( \frac 43 + \frac{42E}{25} \right)\cdot \frac{x_j}{p_j}$.
\end{lemma}

Next, we prove a generalization of Lemma~\ref{lem:discrete-improvement-of-F-with-time},
which works for all Fisher market $\calM$ as described in Theorem~\ref{thm:full-CES}.
The generalization is identical to Lemma~\ref{lem:discrete-improvement-of-F-with-time},
except for replacing the requirement $\lambda \le 1/10$ by $\lambda \le 1/(10E)$.

\paragraph{Proof of the generalization of Lemma~\ref{lem:discrete-improvement-of-F-with-time}}
By having $\lambda \le 1/(10E)$, we ensure that
$$\Gjt ~=~ \frac{\max\{1,\tz_j^t\}}{\lambda p_j^{t-}} ~\ge~ \frac {e^{-E\lambda(\lambda+1)}}2 x_j^{t-} \cdot \frac{10E}{p_j^{t-}} ~\ge~ \frac{4Ex_j^{t-}}{p_j^{t-}},$$
which is greater than or equal to the local parameter $L_j$.
Then by Lemma~\ref{lem:upper-sandwich},
\begin{align*}
&\phi(\bbp^{\alpha(t)}) - \phi(\pt) \\
&~~\ge~~ z_j^t \cdot \Delta p_j^t ~-~ \frac{\Gjt}{2} \cdot (\Delta p_j^t)^2\\
&~~\ge~~ \tz_j^t \cdot \Delta p_j^t - |z_j^t -\tz_j^t| \cdot |\Delta p_j^t| ~-~ \frac{\Gjt}{2} \cdot (\Delta p_j^t)^2\\
&~~=~~ \frac{\Gjt\cdot \Delta p_j^t}{\Dt} \cdot \Delta p_j^t  ~-~ |z_j^t -\tz_j^t| \cdot |\Delta p_j^t| ~-~ \frac{\Gjt}{2} \cdot (\Delta p_j^t)^2\\
&~~\stackrel{(*)}{\ge}~~ \frac{\Gjt\cdot \Delta p_j^t}{\Dt} \cdot \Delta p_j^t ~-~ \frac{\Gjt}{2} \cdot (\Delta p_j^t)^2
~-~ \frac 12 \left( \frac{2}{\Gjt}\cdot (z_j^t -\tz_j^t)^2\cdot \Dt ~+~ \frac {\Gjt}2 \cdot \frac{(\Delta p_j^t)^2}{\Dt} \right)\\
&~~\ge~~ \frac {\Gjt}4 \cdot \frac{(\Delta p_j^t)^2}{\Dt} ~-~ \frac{1}{\Gjt}\cdot (z_j^t -\tz_j^t)^2\cdot \Dt\comm{since $\Dt\le 1$};
\end{align*}
the inequality $(*)$ holds due to the AM-GM inequality.

\paragraph{Proof of Lemma~\ref{lem:W-shift}}
By the elementary inequality $b^2 \ge a^2/2 - (a-b)^2$ for all $a,b\in \rr$, we have
$$
(\Delta p_{j}^t)^2 ~~\ge~~ \frac 12 \cdot (\Delta' p_{j}^t)^2 ~-~ (\Delta p_{j}^t - \Delta' p_{j}^t)^2
~=~ \frac 12 \cdot (\Delta' p_{j}^t)^2 ~-~ \frac{(\Dt)^2}{(\Gjt)^2}\cdot (\tz_j - \tz'_j)^2.$$
The lemma follows on multiplying both sides by $\Gjt / \Dt$.
\section{Leontief Fisher Markets}\label{app:leontief}

It is well-known that Leontief utility functions can be considered as the ``limit'' of CES utility functions as $\rho\ra -\infty$.
We recall that the two properties listed in Section~\ref{sect:lip} and Lemma~\ref{lem:discrete-improvement-of-F-with-time} also hold for Leontief utility functions.
However, we cannot directly apply the analysis for the complementary CES case
to the Leontief Fisher markets for two reasons.
First, while $\phi$ remains convex, it is no longer strongly convex, so $\mu_\phi = 0$.
Second, recall that $\Gktautau ~\ge~ \Theta(1) \cdot \frac{x_{\ktau}(\bbp^{\tau-})}{p_{\ktau}^{\tau-}}$.
For Leontief Fisher markets, it is possible that some good $j$ has zero equilibrium price.
Under this scenario, for convergence to the equilibrium, $\G_j^t$ has to grow towards infinity, and hence $\bG = +\infty$.

Here, we provide additional arguments which build on top of the result that $\Phi(t)$ decreases with $t$, to show that tatonnement
with update rule \eqref{eq:async-tat-CES-rule} still converges toward the market equilibrium.
However, this result does not provide a bound on the rate of convergence.

Tatonnement in Leontief Fisher markets was first analysed by Cheung, Cole and Devanur~\cite{CCD2013}.
They gave a bound on the convergence rate, but with a less natural update rule --- in their update rule,
$\Gjt$ increases with the number of buyers in the market, and is also a function of the demands for all the goods,
both of which seem unnatural, while the $\Gjt$ used here is independent of
the number of buyers and depends only on the demand for good $j$.

\subsection{Analysis}

\begin{lemma}\label{lem:one-big-move-implies-big-progress}
Let $\alpha_j(t),t$ be the times at which two consecutive updates to $p_j$ occur. Let $\Delta t = t - \alpha_j(t)$.
Then $\Phi(\alpha_j(t)) - \Phi(t) ~\ge~ \frac{\Gjt}{8} \cdot \frac{\left(\Delta p_j^t\right)^2}{\Delta t}$.
\end{lemma}

\begin{pf}
First, we use the result in Section~\ref{sect:CES-H-decreasing} to show that $\Phi(\alpha_j(t)) - \Phi(\alpha(t)) \ge 0$.

Then we show that $\Phi(\alpha(t)) - \Phi(t) \ge \frac{\Gjt}{8} \cdot \frac{\left(\Delta p_j^t\right)^2}{\Delta t}$.
To do this, we follow the analysis in Section~\ref{sect:CES-H-decreasing}, except that in~\eqref{eq:decreasing},
we now use the first condition in \eqref{eq:Gamma-requirement} to get the desired improved bound.
\end{pf}

As shown in~\cite{CCD2013}, there exists a finite positive number $U$ which is an upper bound on all the prices
throughout the tatonnement process.

\begin{lemma}\label{lem:big-moves-in-one-time-unit-implies-big-overall-progress}
Suppose that there are consecutive updates to $p_j$ at times $\tau_0 < \tau_1 < \cdots < \tau_m$, where $\tau_m - \tau_0 \leq 2$.
If $\left|p_j^{\tau_0} - p_j^{\tau_m}\right| \ge \epsilon$, where $\epsilon \le 1$, then
$\Phi(\tau_0) - \Phi(\tau_m) ~\ge~ \epsilon^2\cdot \min\left\{\frac{1}{16},\frac{1}{64\lambda U}\right\}$.
\end{lemma}
\begin{pf}
For $q=1,2,\cdots,m$, let $\Delta\pjq$ be the change made to $p_j$ by the update at time $\tau_q$, 
and let $\tzjq$ be the $\tz$-value used for the update,
i.e., $\G_j^{\tau_q} = \frac{\max\{1,\tzjq\}}{\lambda \cdot p_j^{\tau_q^-}}$ and
$\Delta\pjq = \lambda \cdot p_j^{\tau_q^-}\cdot \min\{1,\tzjq\}\cdot \Dt_q$.

If $\tzjq < 1$, then
$$\frac{\G_j^{\tau_q}(\D\pjq)^2}{\Dt_q} = \frac{1}{\lambda \cdot p_j^{\tau_q^-}}\frac{(\D\pjq)^2}{\Dt_q}\geq \frac{1}{\lambda U}\frac{(\D\pjq)^2}{\Dt_q}.$$

If $\tzjq \geq 1$, then
$$\frac{\G_j^{\tau_q}(\D\pjq)^2}{\Dt_q} ~=~ \frac{\tzjq}{\lambda \cdot p_j^{\tau_q^-}}\cdot \lambda^2 \left(p_j^{\tau_q^-}\right)^2\cdot \Dt_q
~=~ \lambda  \cdot  p_j^{\tau_q^-} \tzjq \cdot \Dt_q \geq |\D\pjq|.$$
By Lemma \ref{lem:one-big-move-implies-big-progress},
\begin{align*}
\Phi(\tau_0) - \Phi(\tau_m) ~=~ \sum_{q=1}^m \left(\Phi(\tau_{q-1}) - \Phi(\tau_q)\right) &~\ge~ \frac 18~\sum_{q=1}^m \frac{\G_j^{\tau_q} (\D\pjq)^2}{\Dt_q}\\
&~\ge~ \frac{1}{8\lambda U}\sum_{q:\tzjq < 1} \frac{(\D\pjq)^2}{\Dt_q} ~+~ \frac 18 \sum_{q:\tzjq\geq 1} |\D\pjq|.
\end{align*}

By the assumption $|p_j^{\tau_0} - p_j^{\tau_m}| \geq \epsilon$, $\sum_{q=1}^m |\D\pjq| \geq \epsilon$.
Let $\sigma := \epsilon^{-1} \sum_{q:\tzjq\geq 1} |\D\pjq|$. Then $\sum_{q:\tzjq<1} |\D\pjq| \geq \max\{0,(1-\sigma)\epsilon\}$.
By the Cauchy-Schwarz inequality,
\begin{align*}
\left[\max\{0,(1-\sigma)\epsilon\}\right]^2 \leq \left(\sum_{q:\tzjq<1} |\D\pjq|\right)^2 &~=~ \left(\sum_{q:\tzjq<1} \left|\frac{\D\pjq}{\sqrt{\Dt_q}}\right|\cdot \sqrt{\Dt_q}\right)^2\\
&~\le~ \left(\sum_{q:\tzjq<1} \frac{(\D\pjq)^2}{\Dt_q}\right)\left(\sum_{q:\tzjq<1} \Dt_q\right)\\
&~\le~ 2\sum_{q:\tzjq<1} \frac{(\D\pjq)^2}{\Dt_q},
\end{align*}
as $\tau_m - \tau_0 \leq 2$. Then $\sum_{q:\tzjq<1} \frac{(\D\pjq)^2}{\Dt_q} \geq \frac{1}{2}\left[\max\{0,(1-\sigma)\epsilon\}\right]^2$ and hence
$$\Phi(\tau_0) - \Phi(\tau_m) \geq \frac{1}{16\lambda U} \left[\max\{0,(1-\sigma)\epsilon\}\right]^2 ~+~ \frac{\sigma \epsilon}{8}.$$
By considering the following two cases: $\sigma \ge 1/2$ or $\sigma < 1/2$,
it is easy to show that the minimum value of the RHS of
the above inequality is at least $\epsilon^2\cdot \min\left\{\frac{1}{16},\frac{1}{64\lambda U}\right\}$.
\end{pf}

\begin{corollary}\label{cor:small-move-in-one-time-unit}
For any $\epsilon > 0$, there exists a finite time $T_\epsilon$ such that for any good $j$,
any $t\geq T_\epsilon$, and any $0\leq \Dt \leq 1$, $|p_j^t - p_j^{t+\Dt}| \leq \epsilon$.
\end{corollary}
\begin{pf}
Suppose not, then by Lemma \ref{lem:big-moves-in-one-time-unit-implies-big-overall-progress},
$\Phi$ drops by at least $\epsilon^2\cdot \min\left\{\frac{1}{16},\frac{1}{64\lambda U}\right\}$
infinitely often. But $\Phi(0)$ is finite and $\Phi$ remains positive throughout, a contradiction.
\end{pf}

\bigskip

\begin{pfof}{Theorem \ref{thm:CES-Fisher-cnvge} for the Leontief case}
The proof comprises four steps. We need the following definitions:
for any two price vectors $\bbp^A$ and $\bbp^B$, let $d(\bbp^A,\bbp^B)$ denote the $L_1$ norm distance between the two price vectors,
i.e., $d(\bbp^A,\bbp^B) = \sum_j |p^A_j - p^B_j|$.
For any two sets of price vectors $P^A$ and $P^B$, let
$$d(P^A,P^B) := \inf_{\bbp^A\in P^A,\, \bbp^B\in P^B} d(\bbp^A,\bbp^B).$$

\medskip

\noindent\textbf{Step 1.} Let $\Omega$ be the set of limit points of a tatonnement process. We show that $\Omega$ is non-empty and connected.

\smallskip

Since all prices remain bounded by $U$ throughout the tatonnement process, $\Omega$ is non-empty.

Suppose $\Omega$ is not connected. Let $\Omega_a$ denote a connected component of $\Omega$
that is well separated from $\Omega_b = \Omega\setminus \Omega_a$,
i.e., $d(\Omega_a,\Omega_b) = \ep ' > 0$ (if there is no such $\Omega_a$ then $\Omega$ is connected).
By the definition of limit points, there exists a finite time such that
thereafter the prices in the tatonnement process are always within an $\epsilon'/4$-neighborhood of either $\Omega_a$ or $\Omega_b$.
This forces an infinite number of updates, each separated by at least one time unit,
such that each update makes a change to a price by at least at least $\epsilon'/(2n)$.
This contradicts Corollary \ref{cor:small-move-in-one-time-unit}.

\medskip

\noindent\textbf{Step 2.} Recall that a market equilibrium is a price vector $\bbp^*$ at which for each $j$,
$\ps_j>0$ implies $z_j(\bbp^*)=0$ and $\ps_j=0$ implies $z_j(\bbp^*)\leq 0$.
We define a pseudo-equilibrium: a price vector $\tilde{\bbp}$ is a pseudo-equilibrium if for each $j$,
$\tp_j>0$ implies $z_j(\tilde{\bbp})=0$.
Note that every market equilibrium is a pseudo-equilibrium. We show that all limit points in $\Omega$ are pseudo-equilibria.

\smallskip

\newcommand{\ringp}{\grave{\bbp}}
Suppose not. Let $\bbp'\in \Omega$ be a price vector which is not a pseudo-equilibrium,
i.e., there exists $j$ such that $p'_j>0$ but $z_j(\bbp')\neq 0$.
Let $\ep'' ~\le~ p'_j |z_j(p')|/32$ be a positive number such that for any price vector $\ringp$ in the $\ep''$-neighborhood of $\bbp'$,
we must have $\grave{p}_j\geq p'_j/2$ and $z_j(\ringp)$ lies between $z_j(\bbp')/2$ and $z_j(\bbp')$.

By the definition of limit points, the tatonnement process enters the $(\ep''/2)$-neighborhood of $\bbp'$ infinitely often.
By Corollary \ref{cor:small-move-in-one-time-unit}, there exists a finite time such that subsequently,
every time the tatonnement process enters the $\ep''/2$-neighborhood of $\bbp'$,
it stays in the $\ep''$-neighborhood of $\bbp'$ for at least three time units.
Within the first two time units, $p_j$ is updated at least once,
and by update rule \eqref{eq:async-tat-CES-rule},
such updates will make a total change to $p_j$ of at least $\lambda (p'_j/2) (\left|z_j(\bbp')\right|/2) ~\ge~ 8\ep''$,
which forces quitting the $\ep''$-neighborhood of $p'$ strictly before the three time unit interval, a contradiction.

\medskip

\noindent\textbf{Step 3.} We show that the excess demands at all limit points in $\Omega$ are identical.

\smallskip

For every subset of goods $S$, let $\Omega_S = \{\bbp'\in \Omega\,|\,p'_k > 0 \Leftrightarrow k\in S\}$.
For each buyer, there are two cases:
\begin{itemize}
\item \textbf{the buyer wants at least one good in $S$, say good $\ell$:}\\
Observe that by the definition of pseudo-equilibrium and Step 2, every price vector in $\Omega_S$,
excluding the zero prices in the price vector,
is a market equilibrium for the sub-Leontief-market comprising the goods in $S$.
Codenotti and Varadarajan \cite{CV2004} pointed out that the demands for the goods in $S$ of each buyer are identical at every market equilibrium of the sub-Leontief market,
and hence also in the original Leontief market.
So the buyer demands the same positive but finite amount of good $\ell$ at every price vector in $\Omega_S$
in the original market.
Also note that the buyer always demands the goods in the original market in a fixed proportion.
This forces the demands for the goods not in $S$ of the buyer to also be identical at every price vector in $\Omega_S$.
\item \textbf{the buyer wants no good in $S$:}\\
Then the buyer demands an infinite amount of each good that she wants, and demands zero amount of each good that she does not want.
\end{itemize}
In either case, the buyer's demands for each good at every price vector in $\Omega_S$ are identical,
and hence also the total demand for each good.

Then consider a graph $G$ with each vertex corresponding to a subset of goods $S$ such that $\Omega_S$ is non-empty,
and two vertices $S_1,S_2$ are adjacent if and only if $d\left(\Omega_{S_1},\Omega_{S_2}\right) = 0$.
Since excess demands are a continuous function\footnote{The range of the excess demand functions is the extended real line $\rr \cup \{+\infty\}$; continuity of the excess demand function is w.r.t.~the usual topology on the extended real line. To be specific, if $z_k(p)=+\infty$ for some $p$ and $k$, then for any $M\in \rr$, there exists an $\ep_M>0$ such that $z_k(p)\geq M$ in the $\ep_M$-neighborhood of $p$.} of prices,
if $S_1$ and $S_2$ are adjacent, then the excess demands for all goods at every price vector in $S_1\cup S_2$ are identical.
By Step 1, the graph $G$ is connected, thus the excess demands at all limit points in $\Omega$ are identical.

\bigskip

\noindent\textbf{Step 4.} We show that every limit point in $\Omega$ is indeed a market equilibrium.

\smallskip

Suppose not, i.e., there exists a limit point $\bbp'$ in $\Omega$ which is a pseudo-equilibrium but not a market equilibrium,
i.e., there exists $k$ such that $p'_k=0$ but $z_k(\bbp')>0$.
By Step 3, $z_k$ is positive at every limit point in $\Omega$, and hence every $p_k$ at every limit point must be zero.
By the definition of limit points, for any $\epsilon>0$, beyond a finite time, the tatonnement process must stay within the $\epsilon$-neighborhood of $\Omega$ thereafter.
By choosing a sufficiently small $\epsilon$, $z_k$ is bounded away from zero in the $\epsilon$-neighborhood of $\Omega$, and hence $p_k$ increases indefinitely and eventually $p_k$ becomes so large that the tatonnement process must leave the $\epsilon$-neighborhood of $\Omega$, a contradiction.
\end{pfof}
\end{document}